# Cove-Edged Chiral Graphene Nanoribbons with Chirality-Dependent Bandgap and Carrier Mobility

Kun Liu[1], Wenhao Zheng[2], Silvio Osella[3], Zhen-Lin Qiu[1], Steffen Böckmann[4], Wenhui Niu[1,5], Laura Meingast[6], Hartmut Komber[7], Sebastian Obermann[1], Roland Gillen[6], Mischa Bonn[2], Michael Ryan Hansen[4], Janina Maultzsch[6], Hai I. Wang[2,8], Ji Ma[1,5*], Xinliang Feng[1,5*]

[1] Center for Advancing Electronics Dresden (cfaed) & Faculty of Chemistry and Food Chemistry, Technische Universität Dresden, Mommsenstrasse 4, 01062 Dresden, Germany

[2] Max Planck Institute for Polymer Research, Ackermannweg 10, 55128 Mainz, Germany

[3] Chemical and Biological Systems Simulation Lab, Center of New Technologies, University of Warsaw, Banacha 2C, 02-097 Warsaw, Poland

[4] Institute of Physical Chemistry, Universität Münster, Corrensstraße 28/30, 48149 Münster, Germany

[5] Max Planck Institute of Microstructure Physics, Weinberg 2, Halle, 06120 Germany

[6] Department of Physics, Friedrich-Alexander-Universität Erlangen-Nürnberg (FAU), Staudtstr. 7, 91058, Erlangen, Germany

[7] Leibniz-Institut für Polymerforschung Dresden e. V., Hohe Straße 6, 01069 Dresden, Germany

[8] Nanophotonics, Debye Institute for Nanomaterials Science, Utrecht University, Princetonplein 1, 3584 CC Utrecht, The Netherlands

**ABSTRACT:** Graphene nanoribbons (GNRs) have garnered significant interest due to their highly customizable physicochemical properties and potential utility in nanoelectronics. Besides controlling widths and edge structures, the inclusion of chirality in GNRs brings another dimension for fine-tuning their opto-electronic properties, but related studies remain elusive owing to the absence of feasible synthetic strategies. Here, we demonstrate a novel class of cove-edged chiral GNRs (**CcGNRs**) with a tunable chiral vector (***n*,*m***). Notably, the bandgap and effective mass of **(*n*,2)-CcGNR** show a distinct positive correlation with the increasing value of ***n***, as indicated by theory. Within this GNR family, two representative members, namely **(4,2)-CcGNR** and **(6,2)-CcGNR**, are successfully synthesized. Both **CcGNRs** exhibit prominently curved geometries arising from the incorporated [4]helicene motifs along their peripheries, as also evidenced by the single-crystal structures of two respective model compounds (**1** and **2**). The chemical identities and opto-electronic properties of **(4,2)-** and **(6,2)-CcGNRs** are comprehensively investigated via a combination of IR, Raman, solid-state NMR, UV-vis, and THz spectroscopies as well as theoretical calculations. In line with theoretical expectation, the obtained **(6,2)-CcGNR** possesses a low optical bandgap of 1.37 eV along with charge carrier mobility of ~8 $cm^2 V^{-1} s^{-1}$, whereas the **(4,2)-CcGNR** exhibits a narrower bandgap of 1.26 eV with increased mobility of ~14 $cm^2 V^{-1} s^{-1}$. This work opens up a new avenue to precisely engineer the bandgap and carrier mobility of GNRs by manipulating their chiral vector.

## INTRODUCTION

Graphene nanoribbons (GNRs) possess enormous potential for next-generation semiconductor materials on account of their tunable bandgaps and attractive electronic properties.[1–5] In general, the widths and edge topologies of GNRs exert significant influence over their electronic and magnetic properties.[6,7] Among various methods used to prepare GNRs, bottom-up organic synthesis, utilizing surface-assisted or solution-based chemistry, offers a powerful protocol for fabricating GNRs with structural precision and tailored characteristics.[8–12] Taking advantage of precision organic synthesis, armchair-edged GNRs (**AGNRs**) have been widely studied in the past decade, demonstrating a wide range of bandgaps inversely proportional to the ribbon widths. Moreover, zigzag-edged GNRs (**ZGNRs**) with localized edge states have also been successfully realized via on-surface synthesis under ultrahigh vacuum conditions.[13] Besides, chiral GNRs (**cGNRs**) with combined zigzag and armchair edge orientations represent another important class of GNRs (Figure 1a), which are predicted to possess distinctive properties, such as spin-polarized edge states, chirality-dependent bandgaps, tunable magnetic properties, etc.[14–17] However, experimental access to **cGNRs** remains relatively rare, and there are only a few cases prepared using on-surface synthesis,[18,19] due to the inherent instability associated with the zigzag edges.

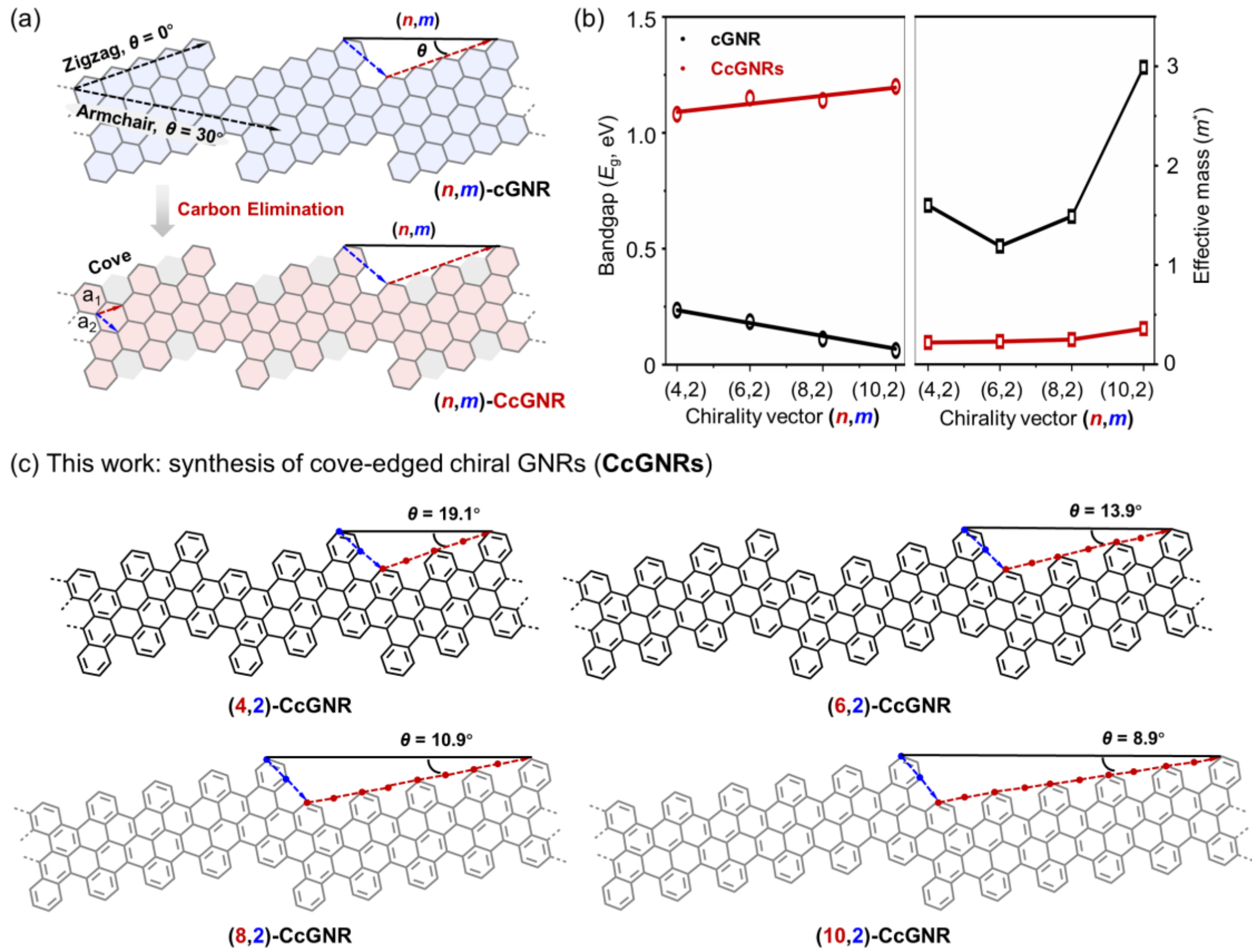

**Figure 1.** (a) Illustration of the pristine ***(n,m)*-cGNRs** and the corresponding cove-edged chiral GNRs (***(n,m)*-CcGNRs**) achieved via periodic carbon elimination. The chirality is described either by the translation vector $\mathbf{C_h}$ defined as $\mathbf{C_h} = \boldsymbol{n}\mathbf{a_1} + \boldsymbol{m}\mathbf{a_2}$, where $n$ and $m$ are lattice translational indices, $\mathbf{a_1}$ and $\mathbf{a_2}$ are the basis vectors, or by the chiral angle $\theta = \arcsin\sqrt{\frac{3}{4}\left(\frac{m^2}{n^2+nm+m^2}\right)}$;[14] (b) Comparison of the calculated bandgap and effective mass between ***(n,m)*-CcGNRs** and ***(n,m)*-cGNRs** (Table S6 and S7); (c) The chemical structures of four family members in ***(n,m)*-CcGNRs**, including the **(4,2)-**, **(6,2)-**, **(8,2)-**, and **(10,2)-CcGNR**, of which **(4,2)-** and **(6,2)-CcGNR** are synthesized in this work. The substituents are omitted for clarity.

From a structural perspective, by periodically eliminating carbon atoms along the zigzag edges of **ZGNR**, the [4]helicene subunits (namely cove edges) are formed on the **ZGNR** peripheries (Figure 1a), giving rise to a family of cove-edged GNRs (**CGNRs**) with nonplanar conformation, good liquid-phase processability, and in some cases, low bandgaps and high charge transport properties.[20–22] Moreover, this "carbon elimination" strategy also has the potential to conquer the stability issue associated with the spin-polarized zigzag edges in GNRs. Inspired by this approach, we envisioned whether it could be applied to the **cGNRs** with the incorporation of cove edges and subsequently control the stability, geometry, and electronic structures of **cGNRs** by varying the chiral vector (Figure 1a).[14,17] To the best of our knowledge, cove-edged chiral GNRs (**CcGNRs**) have yet to be explored both theoretically and experimentally, representing a novel and untapped area of research.

In this work, we report a novel category of chiral GNRs bearing fully cove edges that are realized through the design strategy of "carbon elimination" applied to the backbone of pristine **cGNRs** by manipulating the chiral vector ***(n,m)*** or chiral angle (***θ***) of **cGNRs**. Interestingly, theoretical predictions reveal a distinct trend in the bandgap and effective mass of the resulting ***(n,*2)-CcGNRs** as ***n*** evolves, while pristine ***(n,*2)-cGNRs** display an opposite trend in bandgap evolution and a randomly changing effective mass (Figure 1b). Within this new GNR family, two representative members with the lowest bandgap and effective mass, namely **(4,2)-CcGNR** and **(6,2)-CcGNR**, are successfully synthesized through the Yamamoto polymerization and the subsequent Scholl reaction in solution (Figure 1c). To validate the efficiency of the Scholl reaction, two model compounds **1** and **2** are synthesized as representative fragments of the **(4,2)-CcGNR** and **(6,2)-CcGNR**, respectively (Scheme 1). Crystallographic analysis of **1** and **2** elucidates their characteristic nonplanar alternative up-down topologies attributed to the steric congestion along the cove peripheries. The successful formation of **CcGNRs** is confirmed through a combination of infrared (IR), Raman, and solid-state NMR analysis. Increasing the chiral angle ***θ*** of ***(n,*2)-CcGNRs** is theoretically predicted to reduce the bandgap ($E_g$), which is also validated by experimental comparison of the obtained **(4,2)-CcGNR** ($\theta$ = 19.1°, $E_g$ = 1.26 eV) and **(6,2)-CcGNR** ($\theta$ = 13.9°, $E_g$ = 1.37 eV) through UV-vis absorption spectroscopy. Furthermore, time-resolved terahertz (THz) spectroscopy unveils a charge carrier mobility of ~8.0 $cm^2$ $V^{-1}$ $s^{-1}$ for **(6,2)-CcGNR** in dispersion, which notably increased to ~14 $cm^2$ $V^{-1}$ $s^{-1}$ for **(4,2)-CcGNR**, primarily driven by the enhanced mobility of charge carriers, surpassing that of most solution-synthesized GNRs reported so far.[23–26] Our study offers new insights into the precise customization of GNR's bandgap and carrier mobility via tailoring their chirality.

## RESULTS AND DISCUSSION

**Design Principle Guided by Theoretical Calculation.** As depicted in Figure 1a, the chirality of **cGNR** is described either by the chiral angle ***θ*** or by a chiral vector **(*n*,*m*)**, where ***n*** and ***m*** are the translational indices of the basis vectors of the graphene lattice. Utilizing the same definition, the corresponding **CcGNRs** can be derived through the strategy of periodic carbon elimination along the zigzag edges of **cGNRs**, where the chiral angle ***θ*** gets decreased with increasing ***n*** while keeping ***m*** unchanged. Following this principle, a family of **CcGNRs** with the same unit width is conceived by varying ***n*** and ***m***, in which fully cove-edged **(*n*,2)-CcGNR** is selected as a model to explore the effects of different chirality on their electronic structures. Periodic density functional theory (DFT) calculations are employed to investigate the evolution of bandgaps and effective masses of **(*n*,2)-CcGNRs** by gradually increasing the value of ***n***, in comparison to that of the corresponding **(*n*,2)-cGNRs** (Figure 1b). It is found that varying the value of ***n*** could significantly modulate the electronic structure, affecting both the bandgap and band dispersion of **(*n*,2)-CcGNRs** (Table S6). All of these **CcGNRs** display semiconducting behaviour with narrow bandgaps that increase from 1.08 to 1.21 eV when ***n*** is varied from 4 to 10 (Figure 1c). The most significant impact becomes apparent in their band dispersion as the value of ***n*** varies. Notably, the band dispersion near the valence band maximum (VBM) and the conduction band minimum (CBM) exhibits a near-flat profile for **(8,2)-CcGNR** and **(10,2)-CcGNR**, while a more pronounced dispersion is observed in **(4,2)-CcGNR**, which is directly translated into lower values of reduced masses for both hole and electron carriers ($m^*_h$ and $m^*_e$), thus giving small effective mass ($m^*$) calculated by $\frac{1}{m^*} = \frac{1}{m_e^*} + \frac{1}{m_h^*}$. This characteristic indicates the presence of more extensively delocalized charge carriers in **(4,2)-CcGNR**. As a result, the bandgap and effective mass values exhibit positive correlations with the increase in ***n*** within the **(*n*,2)-CcGNR** system, while that of the pristine **cGNRs** display the opposite trends with semi-metal features (0.06–0.23 eV) and relatively large effective mass (Figure 1b). Motivated by the aforementioned theoretical findings within this **CcGNR** family and taking into account the synthetic feasibility, two representative cases, namely **(4,2)-CcGNR** and **(6,2)-CcGNR**, with narrow bandgaps and low effective masses, were chosen to experimentally investigate the dependency of electronic and opto-electronic properties on the GNR chirality (Figure 1c).

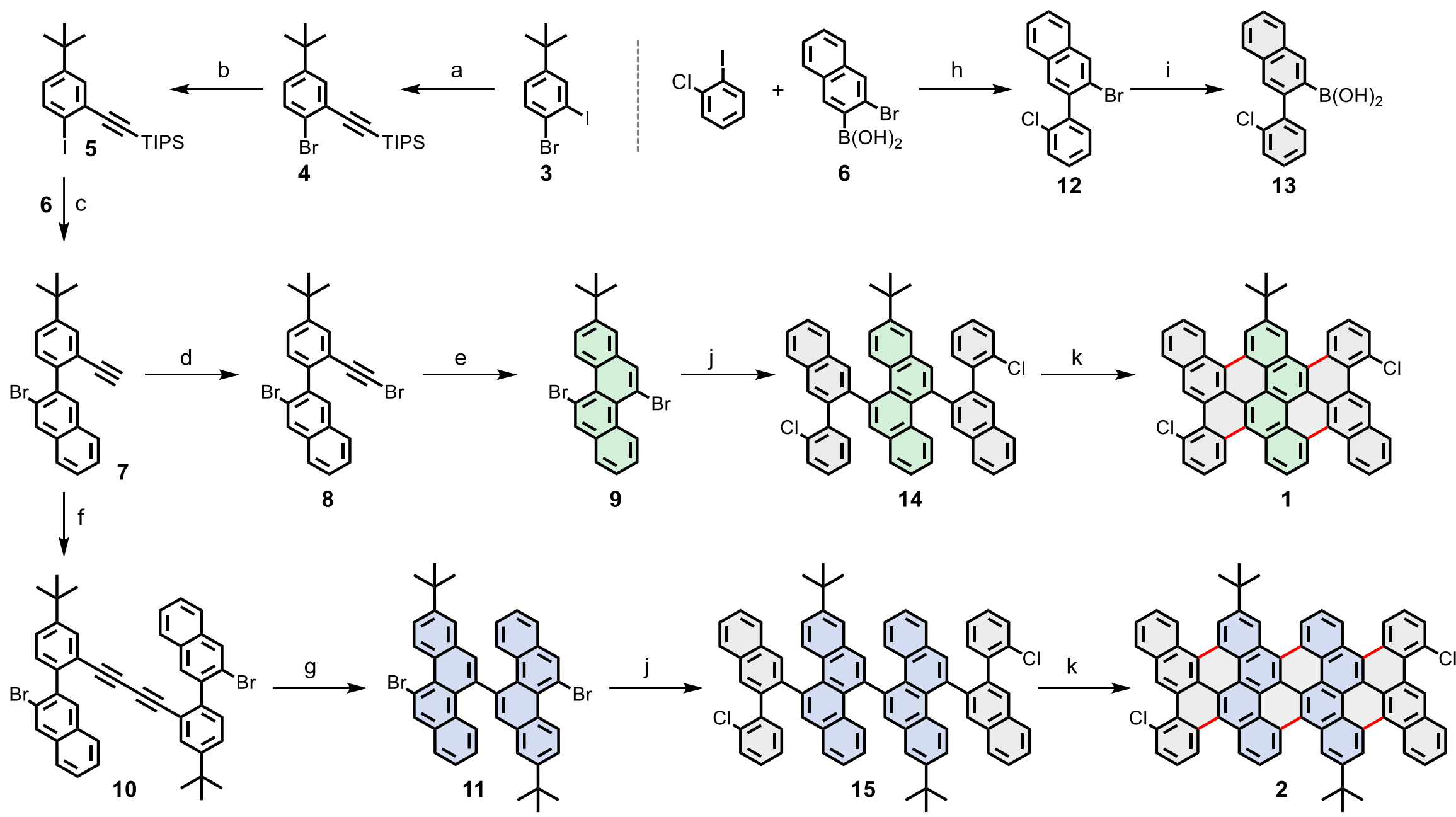

**Scheme 1.** Synthetic route toward model compounds **1** and **2**. Reagents and conditions: (a) (triisopropylsilyl)acetylene, CuI, $PdCl_2(PPh_3)_2$, THF/TEA, r.t., 12 h, 95%; (b) *n*-BuLi, $I_2$, THF, -78 °C, 16 h, 97%; (c) i. **6**, $Pd(PPh_3)_4$, $K_2CO_3$, THF/EtOH/$H_2O$, 60 °C, 36 h; ii. TBAF, THF, r.t., 20 min, 79%; (d) NBS, $AgNO_3$, acetone, r.t., 1 h, 66%; (e) $InCl_3$, toluene, 95 °C, 24 h, 87%; (f) CuI, piperidine, toluene, air, r.t., 6 h, 83%; (g) $PtCl_2$, toluene, 90 °C, 24 h, 93%; (h) $Pd(PPh_3)_4$, $Na_2CO_3$, $n$Bu$_4$NBr, THF/EtOH/$H_2O$, 60 °C, 10 h, 85%; (i) *n*-BuLi, triisopropyl borate, THF, -78 °C, 16 h, 61%; (j) **13**, $Pd(PPh_3)_4$, $K_2CO_3$, toluene/EtOH/$H_2O$, 95 °C, 48 h, 55% for **14** and 60% for **15**; (k) DDQ, TfOH, DCM, 0 °C, 45 min, 62% for **1** and 65% for **2**.

**Synthesis and Characterization of Model Compounds.** As the defined segments of **(4,2)-CcGNR** and **(6,2)-CcGNR**, model compounds **1** and **2** containing chrysene and bischrysene units were synthesized, respectively (Scheme 1). First, ((2-bromo-5-(*tert*-butyl)phenyl)ethynyl)triisopropylsilane (**4**) was prepared in 95% yield by the selective Sonogashira coupling of 1-bromo-4-(*tert*-butyl)-2-iodobenzene (**3**) with the commercially available (triisopropylsilyl)acetylene. Then compound **4** was transformed into ((5-(*tert*-butyl)-2-iodophenyl)ethynyl)triisopropylsilane (**5**) by treatment with *n*-BuLi/iodine in 97% yield. Afterward, a Suzuki coupling of **5** with (3-bromonaphthalen-2-yl)boronic acid (**6**) followed by treatment with tetrabutylammonium fluoride (TBAF) provided 2-bromo-3-(4-(tert-butyl)-2-ethynylphenyl)naphthalene (**7**) with a yield of 79% over two steps. The bromination of the acetylene group in **7** afforded 2-bromo-3-(2-(bromoethynyl)-4-(*tert*-butyl)phenyl)naphthalene (**8**) in 66% yield. Subsequently, 5,11-dibromo-2-(*tert*-butyl)chrysene (**9**) was obtained from **8** in 87% yield via $InCl_3$-catalyzed alkyne benzannulation. Meanwhile, the Glaser coupling of **7** catalyzed by copper iodine (CuI) under air condition gave 1,4-bis(2-(3-bromonaphthalen-2-yl)-5-(*tert*-butyl)phenyl)buta-1,3-diyne (**10**) in 83% yield and the subsequent $PtCl_2$-catalyzed annulation gave the 11,11′-dibromo-5,5′-bischrysene (**11**) in 93% yield. Furthermore, a Suzuki coupling of the commercially available 2-chloroiodobenzene and compound **6** followed by lithiation/borylation gave (3-(2-chlorophenyl)naphthalen-2-yl)boronic acid (**13**) in 52% yield over two steps. After that, the chrysene-based oligomer precursor **14** was synthesized via twofold Suzuki coupling of **9** and **13** in 55% yield. Finally, the Scholl reaction of **14** was carried out by using 2,3-dichloro-5,6-dicyano-1,4-benzoquinone (DDQ)/trifluoromethanesulfonic acid (TfOH), affording the desired model compound **1** in 62% yield. Following the similar synthetic strategy for **1**, model compound **2** was also obtained with a yield of 65% from the bischrysene-based precursor **15** via the same Scholl reaction condition.

The chemical identities of **1** and **2** were first confirmed by MALDI-TOF MS analysis, in which the observed spectra were in good agreement with the simulated isotopic distribution patterns (Figure S2). The chemical structure of **1** was validated by $^1$H NMR spectroscopy and corresponding 2D NMR measurements (Figures S43–S46). However, compound **2** exhibited poor solubility and a strong tendency for aggregation in the common alkyl chloride solvents, resulting in broad $^1$H NMR signals. To our delight, a measurement conducted at 90 °C in the aromatic solvent toluene-$d_8$ gave an evaluable $^1$H NMR spectrum. Single crystals of **1** and **2** were obtained respectively by slow vapor diffusion of methanol into a solution of compound **1** in dichloroethane and a solution of compound **2** in chlorobenzene, which are suitable for single-crystal X-ray diffraction measurements. Compound **1** crystallized in the orthorhombic space group $Pna2_1$, while **2** crystallized in the monoclinic space group *C2*/c (Table S1). As shown in Figures 2a,b, both **1** and **2** possess alternating "up-down" geometries with the large torsion angels within the cove regions (43.23–47.68° for **1** and 38.39–44.99° for **2**) owing to the steric repulsion resulting from the [4]helicene segments at the peripheries. Consequently, elongated C–C bonds (1.42–1.48 Å for **1** and 1.40–1.51 Å for **2**) at the cove edges are afforded, which are comparable with those of the reported cove-edged nanographenes.[27,28] In the solid packing, two enantiomers ((*P*, *M*, *P*, *M*) and (*M*, *P*, *M*, *P*) for **1** and (*P*, *M*, *P*, *P*, *M*, *P*) and (*M*, *P*, *M*, *M*, *P*, *M*) for **2**) are present in a ratio of 1:1 in the unit cell (Figures 2c,d). Notably, the unsymmetrically substituted compound **1** gives rise to racemic dimers via face-to-face $\pi$–$\pi$ interactions with an interlayer distance of 3.7 Å, whereas such dimers further exhibit the herringbone stacking by face-to-edge [C–H···$\pi$] interactions.[29] In contrast, each enantiomer of **2** forms dimers by intermolecular $\pi$–$\pi$ interactions with a similar distance of 3.7 Å, which undergo brick layer stacking along the *b*-axis by aligning with another pair of enantiomeric isomers via $\pi$–$\pi$ interactions.

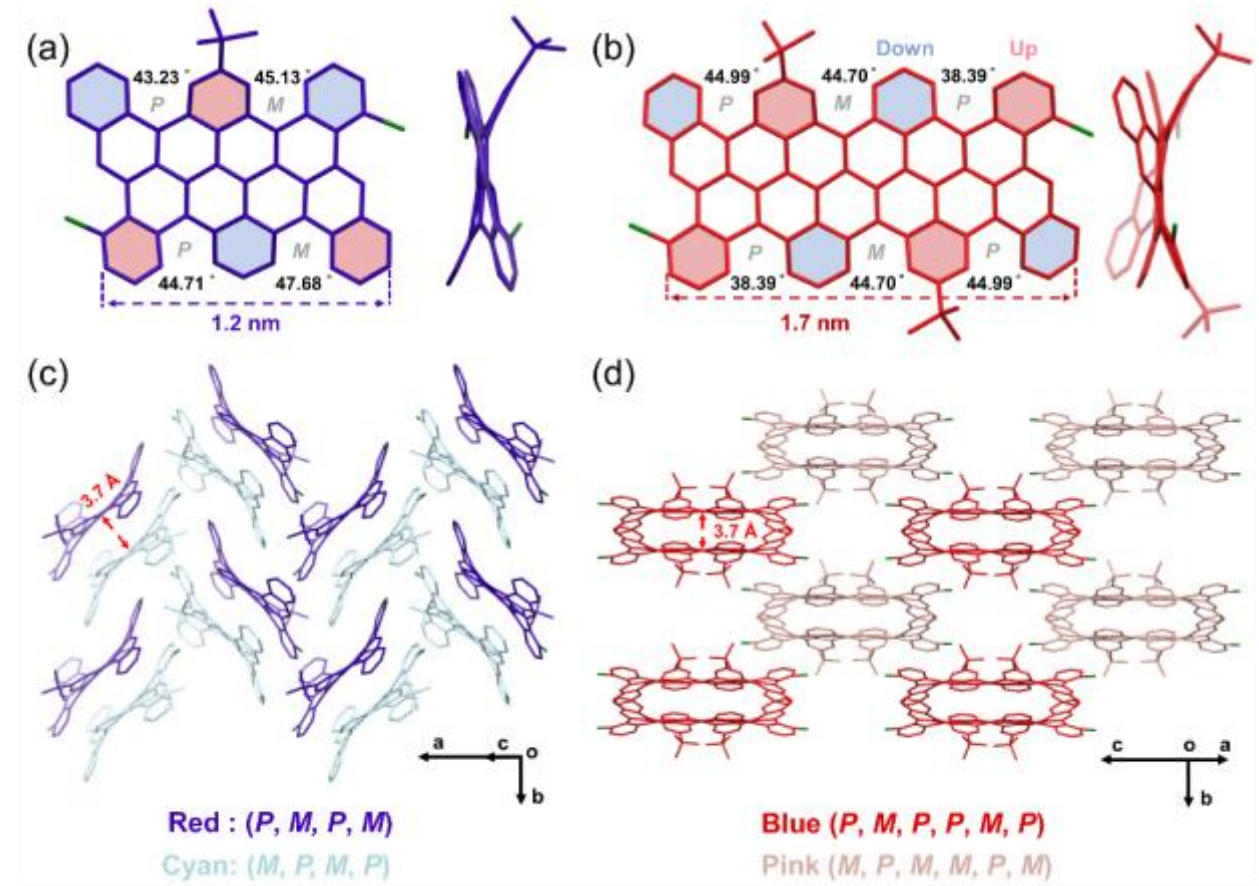

**Figure 2.** X-ray crystallographic structures of **1** and **2**. (a, b) Top and side views of **1** (*P*, *M*, *P*, *M*) and **2** (*P*, *M*, *P*, *P*, *M*, *P*); (c, d) Side view of the crystal packing of **1** and **2**. Hydrogen atoms and solvent molecules are omitted for clarity.

**Synthesis and Characterization of (4,2)-CcGNR and (6,2)-CcGNR.** After confirming the efficient synthesis of model compounds, the synthesis of **(4,2)-CcGNR** and **(6,2)-CcGNR** are further investigated. With the building blocks **14** and **15** in hand, the nickel-mediated Yamamoto polymerization provided the desired polyphenylene precursors **P1** and **P2** in 85% and 89% yields, respectively (Scheme 2). Linear mode MALDI-TOF MS analysis of the crude polymers **P1** and **P2** reveals a family of signals up to *m/z* ~ 9.600 and 10.700, spaced by the repetition unit of 686 and 968 g $mol^{-1}$, respectively (Figure 3a). The obtained crude polymers were then fractionated into three fractions by recycling gel permeation chromatography (GPC), followed by the analysis of analytical size exclusion chromatography (SEC) calibrated by polystyrene (PS) standards. The analysis revealed a close number average molar mass of $M_n$ ~ 10.200 Da with a low polydispersity index (*Đ*) for the first fraction of both polymers with the highest molecular weight ($Đ$=1.06 for **P1**, $Đ$=1.05 for **P2**) (Figures S3 and S4). Finally, oxidative cyclodehydrogenation of the obtained fractions of **P1** and **P2** through the Scholl reaction in DCM with DDQ/TfOH at 0 °C for two days yield the targeted **(4,2)-CcGNR** and **(6,2)-CcGNR** with yields of 86 and 89%, respectively. The average length of the high molecular weight fraction of the resulting **(4,2)-** and **(6,2)-CcGNR** is estimated to be 21 nm, based on the $M_n$ of the corresponding **P1** and **P2** precursors and on the length of the repeat unit estimated from the crystal structures of **1** and **2** (Figures 2a,b). Thanks to their curved geometries and the *tert*-butyl substituents installed on the ribbon edges, both GNRs exhibited good dispersibility in common organic solvents, such as *N*-methyl-2-pyrrolidone (NMP), tetrahydrofuran and 1,2,4-trichlorobenzene.

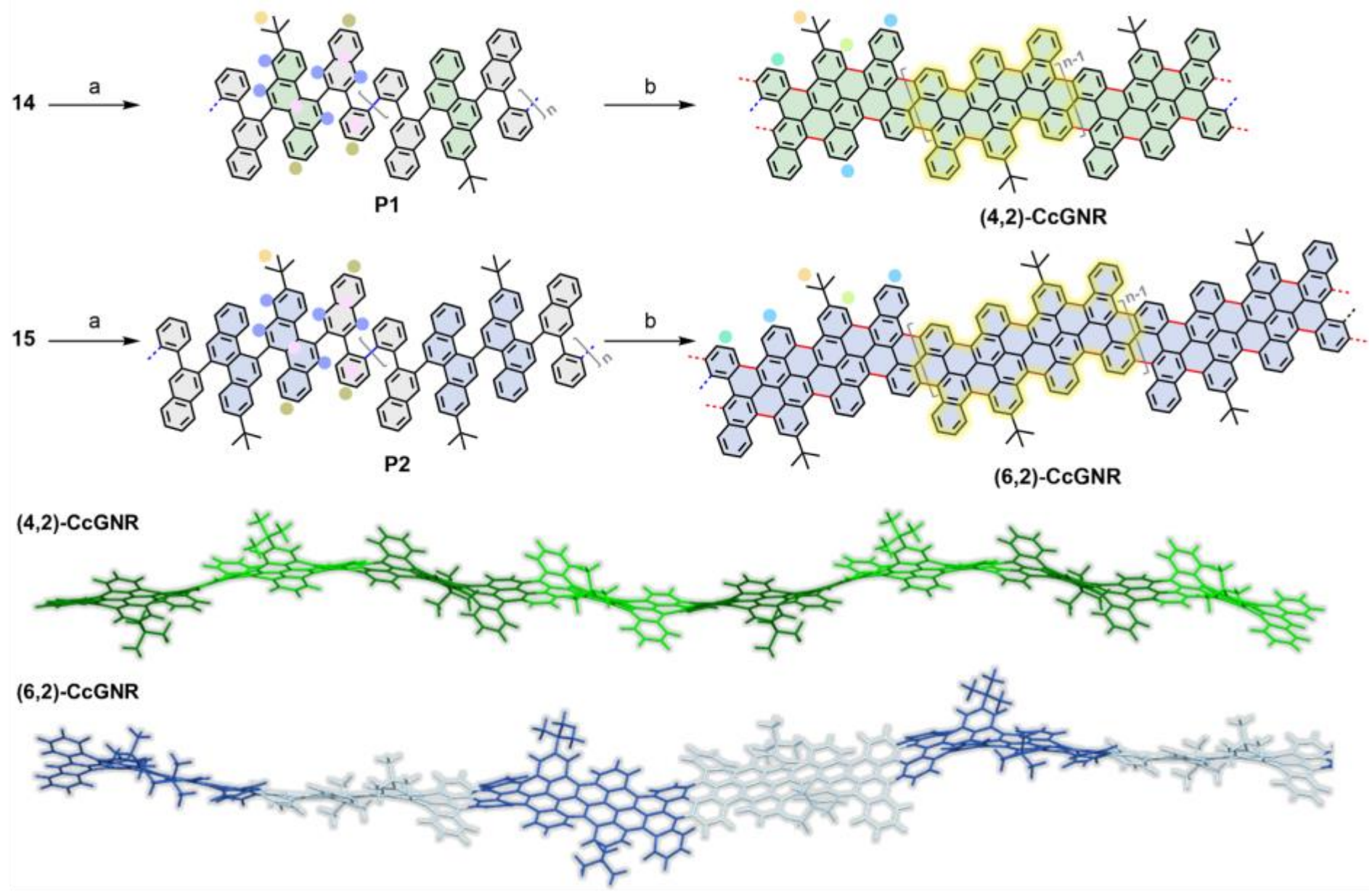

**Scheme 2**. Synthesis of **(4,2)-CcGNR** and **(6,2)-CcGNR** as well as their geometries optimized by DFT simulations. Reagents and conditions: (a) $Ni(COD)_2$, COD, 2,2'-bipyridine, toluene/DMF, 80 °C, 3 days, 85% for **P1** and 89% for **P2**; (b) DDQ, TfOH, DCM, 0 °C to r.t., 2 days, 86% for **(4,2)-CcGNR** and 89% for **(6,2)-CcGNR**. Note that the repeating unit for **(4,2)-** and **(6,2)-CcGNR** is outlined with yellow colors, respectively.

The successful conversion of polymer precursors **P1** and **P2** into **(4,2)-CcGNR** and **(6,2)-CcGNR** was verified through a comprehensive analysis involving IR, Raman, and solid-state NMR spectroscopies. A comparison of the IR spectra of **P1** and **(4,2)-CcGNR** revealed that the C–H stretching vibrations at 3025 and 3053 $cm^{-1}$ in **P1** were diminished after cyclodehydrogenation, whereas a very broad maximum was present close to 3024 $cm^{-1}$ (Figure 3b).[9,30,31] Furthermore, the out-of-plane (*opla*) C–H deformation bands triad at 698 and 744 $cm^{-1}$ (pink and green) originating from disubstituted aromatic rings were attenuated after graphitization, while a strong band associated with TRIO and QUATRO modes (wagging of triply or quadruply adjacent C–H groups, identified by comparison to the HSEH1PBE/6-31G(d) prediction (Table S2–S4)) can be found at 755 $cm^{-1}$ (blue) in **(4,2)-CcGNR**.[31,32] These observations were in line with the variation in the IR spectra of polymer **P2** and **(6,2)-CcGNR** as well, where the C–H stretching region at 3052 $cm^{-1}$, the fingerprint bands at 697 and 744 $cm^{-1}$ were diminished instead of the presence of a joint TRIO and QUATRO mode at 759 $cm^{-1}$. Moreover, Raman spectra of **(4,2)-CcGNR** and **(6,2)-CcGNR** show vibrational modes in the ranges of 150–500 $cm^{-1}$ (inset Figure 3c), 1000–2000 $cm^{-1}$, and 2250–3500 $cm^{-1}$ (Figure 3c), which are confirmed by DFT calculations of the vibrational modes. As the symmetry of the GNR structure is broken by the staircase formation along the nanoribbon axis, there is no symmetry-defined direction of motion of the atoms. Indeed, the calculated vibrations are mixed displacements of different parts of the nanoribbon. In agreement with our experimental results, a characteristic vibrational mode which distinguishes the two nanoribbon structures is not observed in the calculated response. This is probably due to the similarity in the atomic structure of **(4,2)-CcGNR** and **(6,2)-CcGNR**. The DFT simulations revealed that the typical vibrational patterns around 1600 $cm^{-1}$ resemble that of the G mode in graphene/graphite[33,34] (Figure S11a). The lattice vibrations around 1350 $cm^{-1}$ are similar to the vibration leading to the D mode in graphene/graphite[33,34] but include also vibrational contributions of the $CH_3$ groups at the edges (Figure S11b and S11c). Moreover, the solid-state magic-angle spinning (MAS) NMR measurements (Figures S12–14) confirm the successful graphitization of both polymer **P1** and **P2** toward corresponding **(4,2)-CcGNR** and **(6,2)-CcGNR**, in which $^{1}H$ and $^{13}C\{^{1}H\}$ MAS NMR display the broadened signals as a consequence of higher dipolar couplings due to the reduced mobility as well as a broader distribution of isotropic chemical shifts as shown by 2D correlation spectroscopy (Figures S14).

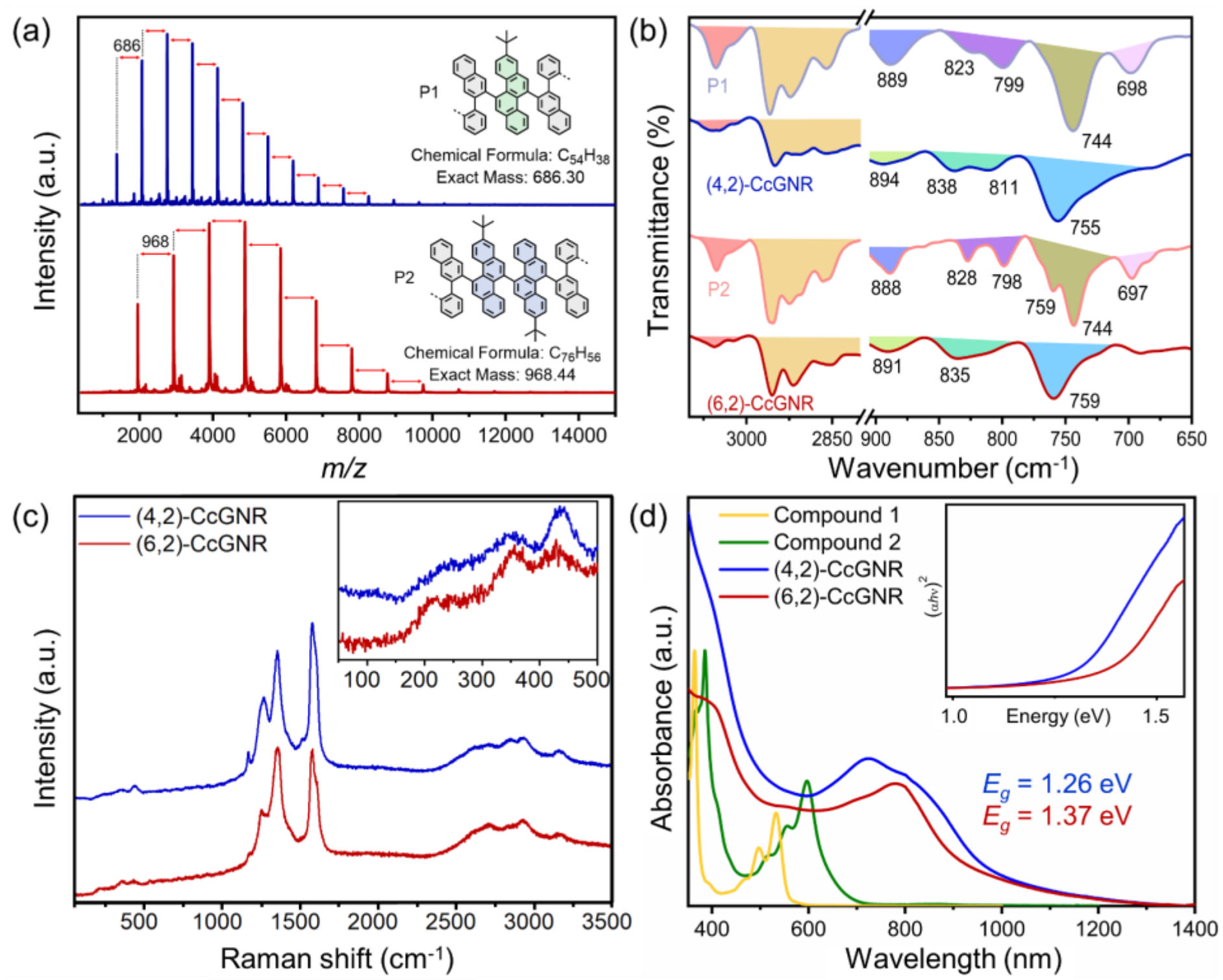

**Figure 3.** Spectroscopic characterizations of **(4,2)-CcGNR** and **(6,2)-CcGNR** and their precursors **P1** and **P2**. (a) MALDI-TOF MS analysis of polymer **P1** and **P2** (matrix: DCTB, linear mode); (b) IR spectra of **P1**, **P2**, **(4,2)-CcGNR** and **(6,2)-CcGNR** with assignments based on DFT simulations. The corresponding vibrational positions are denoted by distinct colored dots in their structures presented in Scheme 2; (c) Raman spectra of **(4,2)-CcGNR** and **(6,2)-CcGNR** recorded with a 532 nm excitation wavelength on a powder sample. The inset shows a zoom into the low-frequency range; (d) UV-vis absorption spectra of model compounds **1** and **2** in $CH_2Cl_2$ ($10^{-5}$ M), as well as **(4,2)-CcGNR** and **(6,2)-CcGNR** in NMP (0.1 mg $mL^{-1}$). The inset shows the corresponding Taut plot of both **CcGNRs**.

**Optical Properties of 1, 2, (4,2)-CcGNR, and (6,2)-CcGNR.** Furthermore, the UV-vis absorption spectra of model compounds **1** and **2** in DCM, as well as the **(4,2)-CcGNR** and **(6,2)-CcGNR** dispersed in NMP, were recorded in Figure 3d. For compound **1**, a distinct absorption peak with the longest wavelength was observed at 533 nm, corresponding to an optical bandgap of 2.25 eV. Compared to that of **1**, compound **2** exhibited an obvious red-shifted absorption peak at 597 nm with an optical gap of 1.98 eV due to its extended π-conjugation. In stark contrast, **(4,2)-CcGNR** displayed considerable bathochromic shift in the absorption toward the NIR region with an absorption maximum at 725 nm and a shoulder peak at 802 nm, while the extended **(6,2)-CcGNR** exhibited an absorption maximum at 781 nm. Based on the Tauc plot method for a direct transition (inset in Figure 3d), the optical bandgap of **(4,2)-CcGNR** was determined to be 1.26 eV, which is approximately 0.11 eV smaller than the optical bandgap (1.37 eV) of **(6,2)-CcGNR**. This result aligns with the trend observed in the DFT-calculated bandgaps (Figure 1b), which also showed a difference of 0.07 eV between **(4,2)-CcGNR** (1.08 eV) and **(6,2)-CcGNR** (1.15 eV) (Table S6).

**Charge Carrier Transport Properties of (4,2)-CcGNR and (6,2)-CcGNR.** We then examined the charge carrier transport characteristics of the obtained GNRs employing ultrafast optical pump-terahertz probe (OPTP) spectroscopy.[35–38] In Figure 4a, we present the time-resolved complex photoconductivity dynamics for both **(4,2)-CcGNR** and **(6,2)-CcGNR** dispersed in 1,2,4-trichlorobenzene, upon optical excitations by ~100 fs laser pulses (with photon energy of 3.10 eV).[39] Note that, compared with **(6,2)-CcGNR**, the elevated real photoconductivity observed in **(4,2)-CcGNR** implies the presence of more mobile charge carriers, in line with the predictions of DFT calculations. To quantify the transport properties of free charge carriers directly after photo-generation, we conducted time-resolved photoconductivity measurement to capture the conductivity response of both GNRs, approximately 1.5 ps after photoexcitation, as depicted in Figure 4b.[35,36] The frequency-resolved complex conductivity can be well-described using the Drude–Smith (DS) model.[36,38,40] This model describes free carrier conductions subject to backscattering effect, due to e.g. structural factors including torsional defects or grain boundaries within the materials. A parameter $c$ is introduced to quantify the backscattering probability, ranging from 0 (isotropic scattering following the Drude model) to -1 (100% backscattering). The fitting yields the charge scattering time $\tau$ of 43 ± 2 and 30 ± 2 fs for **(4,2)-CcGNR** and **(6,2)-CcGNR**, respectively, and a backscattering rate $c$ of ~0.96 for both GNRs.

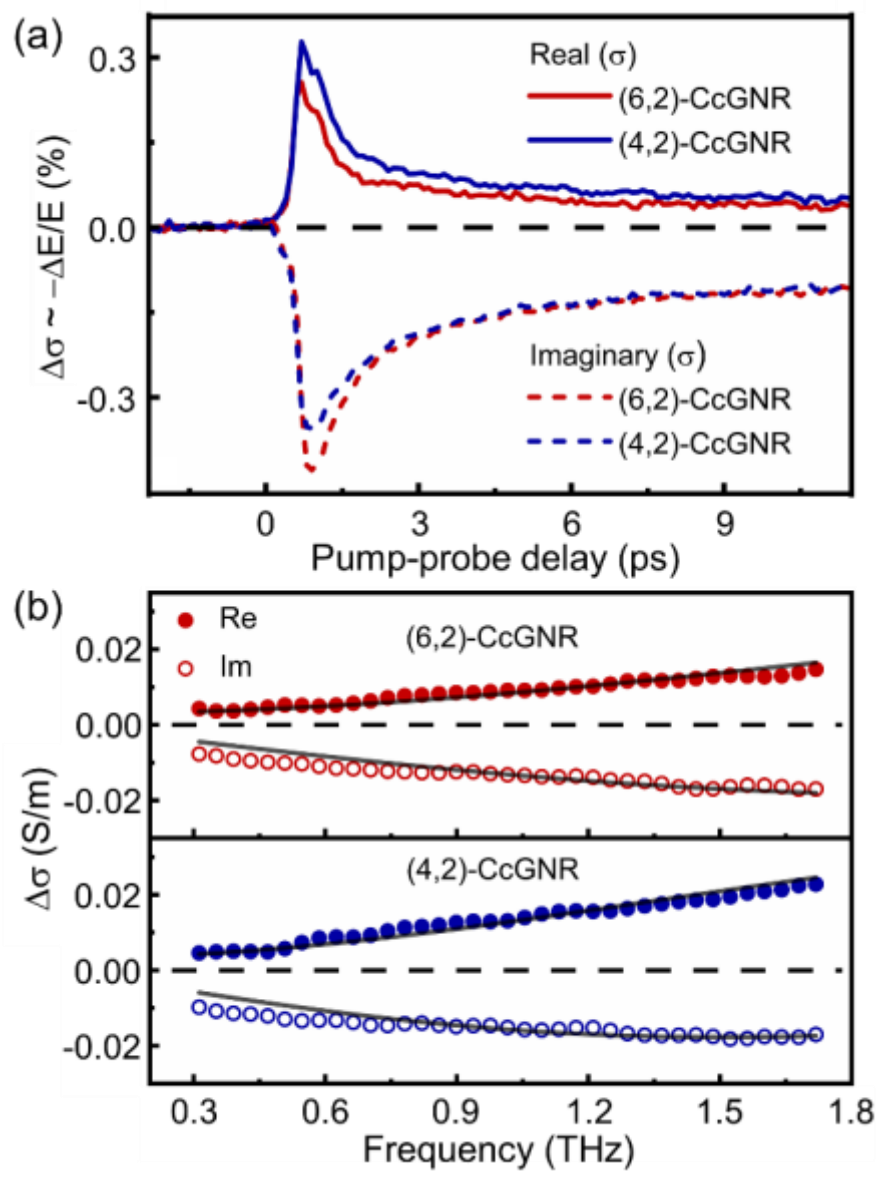

**Figure 4.** (a) Time-resolved complex terahertz photoconductivity of both **(4,2)-CcGNR** and **(6,2)-CcGNR**; (b) Frequency- resolved terahertz conductivity measured at ~1.5 ps after photo-excitation. The black solid lines fit the Drude–Smith model.

Given the ribbon's length of ~20 nm, which is comparable to the mean free propagation path of charge carriers in GNRs (10–20 nm), it is plausible that the backscattering phenomenon in **CcGNR**s primarily stems from scattering events at the ribbon ends.[38,41] Finally, employing the similar effective charge carrier mass $m^*$ obtained by DFT calculations (0.22 $m_0$ and 0.23 $m_0$ for **(4,2)-CcGNR** and **(6,2)-CcGNR** by taking contributions from both electrons and holes (Figure 1b), we estimated the charge mobility in the dc limit $\mu_{dc}$ ($=\frac{e\tau}{m^*}(1+c)$) of 14.2 ± 0.7 and 8.3 ± 0.5 $cm^2$ $V^{-1}$ $s^{-1}$ for **(4,2)-CcGNR** and **(6,2)-CcGNR**, respectively. The relatively high inferred charge carrier mobility in **(4,2)-CcGNR**, compared with that in **(6,2)-CcGNR**, indicates the substantial influence of chirality in dictating charge carrier transport within GNRs.

## CONCLUSION

In summary, we have introduced a novel family of chiral GNRs featured by the cove edge structures, wherein their bandgap and effective mass exhibit chirality-dependent behaviour. Within this family, the successful synthesis of two key members, namely **(4,2)-CcGNR** and **(6,2)-CcGNR** was achieved in solution. Two model compounds **1** and **2** were also synthesized to elucidate the structural features of the corresponding **CcGNRs**, revealing distinct curved geometries resulting from embedded [4]helicene motifs along the edges. The obtained **CcGNRs** were comprehensively characterized by IR, Raman, and solid-state NMR techniques. According to DFT calculations, in contrast to pristine **cGNRs** that have semi-metallic characteristics and large effective masses, **CcGNRs** possess narrow bandgaps and relatively small effective masses that are positively correlated with increasing values of ***n***. Experimentally, both resulting **(4,2)-CcGNR** and **(6,2)-CcGNR** exhibited narrow optical bandgaps of 1.26 and 1.37 eV, respectively, contingent upon the chiral vector (***n*,*m***), in alignment with theoretical predictions. Moreover, the charge carrier mobility underwent a substantial enhancement, increasing from ~8 $cm^2$ $V^{-1}$ $s^{-1}$ for **(6,2)-CcGNR** to ~14 $cm^2$ $V^{-1}$ $s^{-1}$ for **(4,2)-CcGNR**. This study presents a promising avenue for manipulating GNR bandgaps and carrier mobility through chirality modification, thereby greatly expanding the range of potential candidates within the GNR family for integration into nanoelectronics.

## ASSOCIATED CONTENT

**Supporting Information.**

The Supporting Information is available free of charge at http://pubs.acs.org.

Synthetic procedures and characterization data, additional optical spectra of model compounds and polymer precursors, solid-state NMR analysis, DFT calculation details and NMR spectra of new compounds (PDF).

### Corresponding Author

* ji.ma@tu-dresden.de (J. M.)

* xinliang.feng@tu-dresden.de (X. F.)

### Author Contributions

The manuscript was written through contributions of all authors. All authors have given approval to the final version of the manuscript.

### Notes

The authors declare no competing financial interest.

## ACKNOWLEDGMENT

This research was financially supported by the EU Graphene Flagship (Graphene Core 3, 881603), ERC Consolidator Grant (T2DCP, 819698), H2020-MSCA-ITN (ULTIMATE, No. 813036), the Center for Advancing Electronics Dresden (cfaed), and H2020-EU.1.2.2.-FET Proactive Grant (LIGHT-CAP, 101017821). The authors gratefully acknowledge the GWK support for funding this project by providing computing time through the Center for Information Services and HPC (ZIH) at TU Dresden. S.O. thanks the National Science Centre, Poland (grant no. UMO/2020/39/I/ST4/01446). The computation was carried out with the support of the Interdisciplinary Center for Mathematical and Computational Modeling at the University of Warsaw (ICM UW) under grants no. G83-28 and GB80-24. L.M., R.G., J.M. acknowledge support by the Deutsche Forschungsgemeinschaft (DFG, German Research Foundation) – project numbers 447264071 (INST 90/1183-1 FUGG), 182849149 (SFB 953, B13), and 491865171 (GRK 2861). RG, LM and JM also acknowledge the scientific support and HPC resources provided by the Erlangen National High Performance Computing Center (NHR@FAU) of the FAU under the NHR project b181dc. NHR funding is provided by federal and Bavarian state authorities. NHR@FAU hardware is partially funded by the German Research Foundation (DFG) – 440719683.

## REFERENCES

(1) Li, X.; Wang, X.; Zhang, L.; Lee, S.; Dai, H. Chemically Derived, Ultrasmooth Graphene Nanoribbon Semiconductors. *Science* **2008**, *319*, 1229–1232.

(2) Jiao, L.; Zhang, L.; Wang, X.; Diankov, G.; Dai, H. Narrow Graphene Nanoribbons from Carbon Nanotubes. *Nature* **2009**, *458*, 877–880.

(3) Cai, J.; Ruffieux, P.; Jaafar, R.; Bieri, M.; Braun, T.; Blankenburg, S.; Muoth, M.; Seitsonen, A. P.; Saleh, M.; Feng, X.; Müllen, K.; Fasel, R. Atomically Precise Bottom-up Fabrication of Graphene Nanoribbons. *Nature* **2010**, *466*, 470–473.

(4) Narita, A.; Wang, X.-Y.; Feng, X.; Müllen, K. New Advances in Nanographene Chemistry. *Chem. Soc. Rev.* **2015**, *44*, 6616–6643.

(5) Magda, G. Z.; Jin, X.; Hagymási, I.; Vancsó, P.; Osváth, Z.; Nemes-Incze, P.; Hwang, C.; Biró, L. P.; Tapasztó, L. Room-Temperature Magnetic Order on Zigzag Edges of Narrow Graphene Nanoribbons. *Nature* **2014**, *514*, 608–611.

(6) Houtsma, R. S. K.; de la Rie, J.; Stöhr, M. Atomically Precise Graphene Nanoribbons: Interplay of Structural and Electronic Properties. *Chem. Soc. Rev.* **2021**, *50*, 6541–6568.

(7) Nakada, K.; Fujita, M.; Dresselhaus, G.; Dresselhaus, M. S. Edge State in Graphene Ribbons: Nanometer Size Effect and Edge Shape Dependence. *Phys. Rev. B* **1996**, *54*, 17954–17961.

(8) Yang, X.; Dou, X.; Rouhanipour, A.; Zhi, L.; Räder, H. J.; Müllen, K. Two-Dimensional Graphene Nanoribbons. *J. Am. Chem. Soc.* **2008**, *130*, 4216–4217.

(9) Narita, A.; Feng, X.; Hernandez, Y.; Jensen, S. A.; Bonn, M.; Yang, H.; Verzhbitskiy, I. A.; Casiraghi, C.; Hansen, M. R.; Koch, A. H. R.; Fytas, G.; Ivasenko, O.; Li, B.; Mali, K. S.; Balandina, T.; Mahesh, S.; De Feyter, S.; Müllen, K. Synthesis of Structurally Well-Defined and Liquid-Phase-Processable Graphene Nanoribbons. *Nat. Chem.* **2014**, *6*, 126–132.

(10) Chen, L.; Hernandez, Y.; Feng, X.; Müllen, K. From Nanographene and Graphene Nanoribbons to Graphene Sheets: Chemical Synthesis. *Angew. Chem. Int. Ed.* **2012**, *51*, 7640–7654.

(11) Gao, Y.; Huang, L.; Cao, Y.; Richter, M.; Qi, J.; Zheng, Q.; Yang, H.; Ma, J.; Chang, X.; Fu, X.; Palma, C.-A.; Lu, H.; Zhang, Y.-Y.; Cheng, Z.; Lin, X.; Ouyang, M.; Feng, X.; Du, S.; Gao, H.-J. Selective Activation of Four Quasi-Equivalent C–H Bonds Yields N-Doped Graphene Nanoribbons with Partial Corannulene Motifs. *Nat. Commun.* **2022**, *13*, 6146.

(12) Zhang, J.-J.; Liu, K.; Xiao, Y.; Yu, X.; Huang, L.; Gao, H.-J.; Ma, J.; Feng, X. Precision Graphene Nanoribbon Heterojunctions by Chain-Growth Polymerization. *Angew. Chem. Int. Ed.* **2023**, *62*, e202310880.

(13) Ruffieux, P.; Wang, S.; Yang, B.; Sánchez-Sánchez, C.; Liu, J.; Dienel, T.; Talirz, L.; Shinde, P.; Pignedoli, C. A.; Passerone, D.; Dumslaff, T.; Feng, X.; Müllen, K.; Fasel, R. On-Surface Synthesis of Graphene Nanoribbons with Zigzag Edge Topology. *Nature* **2016**, *531*, 489–492.

(14) Golor, M.; Lang, T. C.; Wessel, S. Quantum Monte Carlo Studies of Edge Magnetism in Chiral Graphene Nanoribbons. *Phys. Rev. B* **2013**, *87*, 155441.

(15) Sánchez-Sánchez, C.; Dienel, T.; Deniz, O.; Ruffieux, P.; Berger, R.; Feng, X.; Müllen, K.; Fasel, R. Purely Armchair or Partially Chiral: Noncontact Atomic Force Microscopy Characterization of Dibromo-Bianthryl-Based Graphene Nanoribbons Grown on Cu(111). *ACS Nano* **2016**, *10*, 8006–8011.

(16) Wang, X.-Y.; Urgel, J. I.; Barin, G. B.; Eimre, K.; Di Giovannantonio, M.; Milani, A.; Tommasini, M.; Pignedoli, C. A.; Ruffieux, P.; Feng, X.; Fasel, R.; Müllen, K.; Narita, A. Bottom-Up Synthesis of Heteroatom-Doped Chiral Graphene Nanoribbons. *J. Am. Chem. Soc.* **2018**, *140*, 9104–9107.

(17) Jiang, Z.; Song, Y. Band Gap Oscillation and Novel Transport Property in Ultrathin Chiral Graphene Nanoribbons. *Phys. B Condens. Matter* **2015**, *464*, 61–67.

(18) Lawrence, J.; Berdonces-Layunta, A.; Edalatmanesh, S.; Castro-Esteban, J.; Wang, T.; Jimenez-Martin, A.; de la Torre, B.; Castrillo-Bodero, R.; Angulo-Portugal, P.; Mohammed, M. S. G.; Matěj, A.; Vilas-Varela, M.; Schiller, F.; Corso, M.; Jelinek, P.; Peña, D.; de Oteyza, D. G. Circumventing the Stability Problems of Graphene Nanoribbon Zigzag Edges. *Nat. Chem.* **2022**, *14*, 1451–1458.

(19) Li, J.; Sanz, S.; Merino-Díez, N.; Vilas-Varela, M.; Garcia-Lekue, A.; Corso, M.; de Oteyza, D. G.; Frederiksen, T.; Peña, D.; Pascual, J. I. Topological Phase Transition in Chiral Graphene Nanoribbons: From Edge Bands to End States. *Nat. Commun.* **2021**, *12*, 5538.

(20) Niu, W.; Ma, J.; Soltani, P.; Zheng, W.; Liu, F.; Popov, A. A.; Weigand, J. J.; Komber, H.; Poliani, E.; Casiraghi, C.; Droste, J.; Hansen, M. R.; Osella, S.; Beljonne, D.; Bonn, M.; Wang, H. I.; Feng, X.; Liu, J.; Mai, Y. A Curved Graphene Nanoribbon with Multi-Edge Structure and High Intrinsic Charge Carrier Mobility. *J. Am. Chem. Soc.* **2020**, *142*, 18293–18298.

(21) Wang, X.; Ma, J.; Zheng, W.; Osella, S.; Arisnabarreta, N.; Droste, J.; Serra, G.; Ivasenko, O.; Lucotti, A.; Beljonne, D.; Bonn, M.; Liu, X.; Hansen, M. R.; Tommasini, M.; De Feyter, S.; Liu, J.; Wang, H. I.; Feng, X. Cove-Edged Graphene Nanoribbons with Incorporation of Periodic Zigzag-Edge Segments. *J. Am. Chem. Soc.* **2022**, *144*, 228–235.

(22) Dubey, R. K.; Melle-Franco, M.; Mateo-Alonso, A. Twisted Molecular Nanoribbons with up to 53 Linearly-Fused Rings. *J. Am. Chem. Soc.* **2021**, *143*, 6593–6600.

(23) Niu, W.; Ma, J.; Feng, X. Precise Structural Regulation and Band-Gap Engineering of Curved Graphene Nanoribbons. *Acc. Chem. Res.* **2022**, *55*, 3322–3333.

(24) Obermann, S.; Zheng, W.; Melidonie, J.; Böckmann, S.; Osella, S.; Arisnabarreta, N.; Guerrero-León, L. A.; Hennersdorf, F.; Beljonne, D.; Weigand, J. J.; Bonn, M.; De Feyter, S.; Hansen, M. R.; Wang, H. I.; Ma, J.; Feng, X. Curved Graphene Nanoribbons Derived from Tetrahydropyrene-Based Polyphenylenes *via* One-Pot K-Region Oxidation and Scholl Cyclization. *Chem. Sci.* **2023**, *14*, 8607–8614.

(25) Yao, X.; Zheng, W.; Osella, S.; Qiu, Z.; Fu, S.; Schollmeyer, D.; Müller, B.; Beljonne, D.; Bonn, M.; Wang, H. I.; Müllen, K.; Narita, A. Synthesis of Nonplanar Graphene Nanoribbon with Fjord Edges. *J. Am. Chem. Soc.* **2021**, *143*, 5654–5658.

(26) Yang, L.; Ma, J.; Zheng, W.; Osella, S.; Droste, J.; Komber, H.; Liu, K.; Böckmann, S.; Beljonne, D.; Hansen, M. R.; Bonn, M.; Wang, H. I.; Liu, J.; Feng, X. Solution Synthesis and Characterization of a Long and Curved Graphene Nanoribbon with Hybrid Cove–Armchair–Gulf Edge Structures. *Adv. Sci.* **2022**, *9*, 2200708.

(27) Gu, Y.; Vega-Mayoral, V.; Garcia-Orrit, S.; Schollmeyer, D.; Narita, A.; Cabanillas-González, J.; Qiu, Z.; Müllen, K. Cove-Edged Hexa-Peri-Hexabenzo-Bis-Peri-Octacene: Molecular Conformations and Amplified Spontaneous Emission. *Angew. Chem. Int. Ed.* **2022**, *61*, e202201088.

(28) Gu, Y.; Muñoz-Mármol, R.; Wu, S.; Han, Y.; Ni, Y.; Díaz-García, M. A.; Casado, J.; Wu, J. Cove-Edged Nanographenes with Localized Double Bonds. *Angew. Chem.* **2020**, *132*, 8190–8194.

(29) Yao, Z.-F.; Wang, J.-Y.; Pei, J. Control of π–π Stacking via Crystal Engineering in Organic Conjugated Small Molecule Crystals. *Cryst. Growth Des.* **2018**, *18*, 7–15.

(30) Centrone, A.; Brambilla, L.; Renouard, T.; Gherghel, L.; Mathis, C.; Müllen, K.; Zerbi, G. Structure of New Carbonaceous Materials: The Role of Vibrational Spectroscopy. *Carbon* **2005**, *43*, 1593–1609.

(31) Hu, Y.; Xie, P.; De Corato, M.; Ruini, A.; Zhao, S.; Meggendorfer, F.; Straasø, L. A.; Rondin, L.; Simon, P.; Li, J.; Finley, J. J.; Hansen, M. R.; Lauret, J.-S.; Molinari, E.; Feng, X.; Barth, J. V.; Palma, C.-A.; Prezzi, D.; Müllen, K.; Narita, A. Bandgap Engineering of Graphene Nanoribbons by Control over Structural Distortion. *J. Am. Chem. Soc.* **2018**, *140*, 7803–7809.

(32) Schwab, M. G.; Narita, A.; Hernandez, Y.; Balandina, T.; Mali, K. S.; De Feyter, S.; Feng, X.; Müllen, K. Structurally Defined Graphene Nanoribbons with High Lateral Extension. *J. Am. Chem. Soc.* **2012**, *134*, 18169–18172.

(33) Jorio, A.; Saito, R.; Dresselhaus, G.; Dresselhaus, M. S. *Raman Spectroscopy in Graphene Related Systems*, 1st ed.; WILEY-VCH Verlag GmbH & Co. KGaA, 2011.

(34) Mohr, M.; Maultzsch, J.; Dobardžić, E.; Reich, S.; Milošević, I.; Damnjanović, M.; Bosak, A.; Krisch, M.; Thomsen, C. Phonon Dispersion of Graphite by Inelastic X-Ray Scattering. *Phys. Rev. B* **2007**, *76*, 035439.

(35) Zheng, W.; Zorn, N. F.; Bonn, M.; Zaumseil, J.; Wang, H. I. Probing Carrier Dynamics in Sp3-Functionalized Single-Walled Carbon Nanotubes with Time-Resolved Terahertz Spectroscopy. *ACS Nano* **2022**, *16*, 9401–9409.

(36) Ulbricht, R.; Hendry, E.; Shan, J.; Heinz, T. F.; Bonn, M. Carrier Dynamics in Semiconductors Studied with Time-Resolved Terahertz Spectroscopy. *Rev. Mod. Phys.* **2011**, *83*, 543–586.

(37) Zheng, W.; Sun, B.; Li, D.; Gali, S. M.; Zhang, H.; Fu, S.; Di Virgilio, L.; Li, Z.; Yang, S.; Zhou, S.; Beljonne, D.; Yu, M.; Feng, X.; Wang, H. I.; Bonn, M. Band Transport by Large Fröhlich Polarons in MXenes. *Nat. Phys.* **2022**, *18*, 544–550.

(38) Jensen, S. A.; Ulbricht, R.; Narita, A.; Feng, X.; Müllen, K.; Hertel, T.; Turchinovich, D.; Bonn, M. Ultrafast Photoconductivity of Graphene Nanoribbons and Carbon Nanotubes. *Nano Lett.* **2013**, *13*, 5925–5930.

(39) Tries, A.; Osella, S.; Zhang, P.; Xu, F.; Ramanan, C.; Kläui, M.; Mai, Y.; Beljonne, D.; Wang, H. I. Experimental Observation of Strong Exciton Effects in Graphene Nanoribbons. *Nano Lett.* **2020**, *20*, 2993–3002.

(40) Cocker, T. L.; Baillie, D.; Buruma, M.; Titova, L. V.; Sydora, R. D.; Marsiglio, F.; Hegmann, F. A. Microscopic Origin of the Drude-Smith Model. *Phys. Rev. B* **2017**, *96*, 205439.

(41) Dubey, R. K.; Marongiu, M.; Fu, S.; Wen, G.; Bonn, M.; Wang, H. I.; Melle-Franco, M.; Mateo-Alonso, A. Accelerated Iterative Synthesis of Ultralong Graphene Nanoribbons with Full Atomic Precision. *Chem* **2023**, *9*, 1–14.

Table of Contents

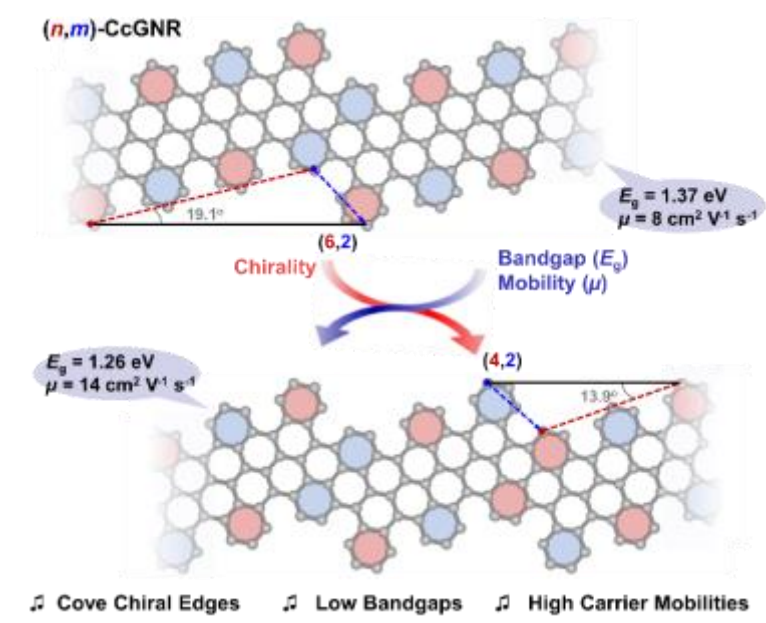

(n,m)-CcGNR
19.1°
(6,2)
Eg = 1.37 eV
μ = 8 cm² V-1 s-1
Chirality
Bandgap (Eg)
Mobility (μ)
Eg = 1.26 eV
μ = 14 cm² V-1 s-1
(4,2)
13.9°
♫ Cove Chiral Edges
♫ Low Bandgaps
♫ High Carrier Mobilities